\documentclass[11pt,a4paper]{article}

\pdfoutput=1
\usepackage{jcappub}
\usepackage{slashed}

\usepackage{multirow}
\usepackage{amssymb}
\usepackage{ulem}
\usepackage{cancel}
\usepackage{verbatim}
\usepackage{caption}
\usepackage{subcaption}
\usepackage[utf8]{inputenc}
\usepackage{xcolor}

\usepackage{array}
\newcolumntype{L}[1]{>{\raggedright\let\newline\\\arraybackslash\hspace{0pt}}m{#1}}
\newcolumntype{C}[1]{>{\centering\let\newline\\\arraybackslash\hspace{0pt}}m{#1}}
\newcolumntype{R}[1]{>{\raggedleft\let\newline\\\arraybackslash\hspace{0pt}}m{#1}}

\allowdisplaybreaks

\definecolor{darkred}{rgb}{0.5,0,0}
\definecolor{darkgreen}{rgb}{0,0.5,0}
\definecolor{darkblue}{rgb}{0,0,0.5}

\usepackage{hyperref}
\hypersetup{ colorlinks,
linkcolor=darkblue,
filecolor=darkgreen,
urlcolor=darkred,
citecolor=darkblue }

\newcommand{\inspire}[1]{\href{http://inspirehep.net/search?p=find+J+#1}
 {{\color{black}[{\color{blue} {\small in}SPIRE}]}}}
\newcommand{\book}[1]{\href{http://inspirehep.net/search?p=#1}
 {{\color{black}[{\color{blue} {\small in}SPIRE}]}}}

\newcommand{\inspired}[1]{\href{http://inspirehep.net/search?p=#1}
 {{\color{black}[{\color{blue} {\small in}SPIRE}]}}}

\begin{document}

%\begin{flushright}
%{\large \tt 
%TTK-12-15}
%\end{flushright}

\title{How to relax the cosmological neutrino mass bound}

\date{\today}

\author[a]{Isabel M.~Oldengott,}
\author[a]{Gabriela Barenboim,}
\author[b]{Sarah Kahlen,}
\author[c]{Jordi Salvado,}
\author[d]{Dominik J.~Schwarz}

\emailAdd{isabel.oldengott@uv.es}

\affiliation[a]{Departament de Física Teórica and IFIC, Universitat de València-CSIC, E-46100, Burjassot, Spain}
\affiliation[b]{Institut für Physik, Universität Oldenburg, Postfach 2503, D-26111
Oldenburg, Germany}
\affiliation[c]{Department de Fisíca Quàntica i Astrofísica and Institut de Ciencies del Cosmos, Univer-
sitat de Barcelona, Diagonal 647, E-08028 Barcelona, Spain}
\affiliation[d]{Fakultät für Physik, Universität Bielefeld, Postfach 100131, 33501 Bielefeld, Germany}

\abstract{We study the impact of non-standard  momentum distributions of cosmic neutrinos on the anisotropy spectrum of the cosmic microwave background and the matter power spectrum of the large scale structure. We show that the neutrino distribution has almost no \textit{unique} observable imprint, as it is almost entirely degenerate with the effective number of neutrino flavours, $N_{\mathrm{eff}}$, \textit{and} the neutrino mass, $m_{\nu}$. Performing a Markov chain Monte Carlo analysis with current cosmological data, we demonstrate that the neutrino mass bound heavily depends on the assumed momentum distribution of relic neutrinos. The message of this work is simple and has to our knowledge not been pointed out clearly before: Cosmology allows that neutrinos have larger masses if their average momentum is larger than that of a perfectly thermal distribution. Here we provide an example in which the mass limits are relaxed by a factor of two. }

\maketitle   

\flushbottom
%%%%%%%%%%%%%%%%%%%%%%%%%%%%%%%%%%%%%%%%%%%%%%%%%%%%%%%%

\section{Introduction}
The neutrino sector is one of the most obscure sectors within the standard model of particle physics. The nature of neutrinos, the neutrino mass production mechanism and even the values of the neutrino masses are still open puzzles and matter to an active field of research.

Since the observation of neutrino oscillations revealed that neutrinos are massive, the determination of the neutrino masses has become the holy grail of neutrino physics. While the neutrino mass splittings have already been measured \cite{Esteban:2016qun,deSalas:2017kay}, the neutrino mass ordering as well as the mass of the lightest mass state are still undetermined. 
The strongest bound on the neutrino mass from laboratory searches currently comes from the Troitzk and Mainz experiments, namely $m_{\nu_e}<2.05$ eV (95\% CL) \cite{Aseev:2011dq} and $m_{\nu_e}< 2.3$ eV (95\% CL) \cite{Kraus:2004zw}, where $m_{\nu_e} = (\sum_i |U_{e i}|^2 m_i^2 )^{1/2}$ and $U_{e i}$ are the entries of the Pontecorvo–Maki–Nakagawa–Sakata matrix. The tritium decay experiment KATRIN aims to reach a sensitivity of $m_{\nu_e}<$0.2 eV at 90\% CL \cite{Osipowicz:2001sq}. 

The by far strongest constraint on the sum of the neutrino masses today however comes from cosmological observations. Measurements of the temperature and polarization anisotropy spectrum of the cosmic microwave background (CMB) with the Planck satellite in combination with other cosmological measurements like baryonic acoustic oscillations (BAO) show that $\sum m_{\nu}< 0.12$ eV (95\% CL) \cite{Aghanim:2018eyx}. This tight constraint has also been derived in the analysis of \cite{Vagnozzi:2017ovm} including galaxy power spectra but with Planck 2015 data \cite{Ade:2015xua}. Future data analyses promise even tighter bounds and will be realized in the next decade \cite{Archidiacono:2016lnv,Sprenger:2018tdb,Brinckmann:2018owf,Boyle:2017lzt,Allison:2015qca,Mishra-Sharma:2018ykh,Yu:2018tem}. The cosmological mass bounds relax to some extent when marginalized over different extensions of the standard $\Lambda$CDM model \cite{Gariazzo:2018meg} but were found to be relatively robust \cite{Ade:2015xua,DiValentino:2015ola}. 

While most of the $\Lambda$CDM extensions studied in 
\cite{Ade:2015xua,Brinckmann:2018owf,DiValentino:2015ola,Archidiacono:2016lnv,Mishra-Sharma:2018ykh,Boyle:2017lzt,Gariazzo:2018meg}
 are addressing non-neutrino properties, we show how the cosmological neutrino mass bound can be lowered by directly targeting an assumption that we usually make about cosmological neutrinos: We abandon the assumption of a Fermi-Dirac distribution for neutrinos and allow non-thermal features in the neutrino phase-space distribution function. The idea behind is as simple as follows: By allowing relic neutrinos to have a higher average-momentum they can have much larger masses. 

We also address the question which impact the neutrino distribution function has in the first place on the CMB  and the large scale structure (LSS) of the Universe. We show that there is almost no \textit{unique} feature of the neutrino distribution function in the CMB and LSS since non-thermal distortions can almost be entirely mimicked by changes in the values of the effective number of neutrino flavours $N_{\text{eff}}$ \textit{and} $\sum m_{\nu}$. 

While the degeneracy between the neutrino distribution function and $N_{\text{eff}}$ is widely known, its intrinsic degeneracy with $\sum m_{\nu}$ has been much less communicated. The latter has however already been pointed out in \cite{Cuoco:2005qr}. But at the time of this work cosmological data were still much less precise than today. In particular at the time of \cite{Cuoco:2005qr,Crotty:2004gm,Hannestad:2003ye}, there was still a large degeneracy between $\sum m_{\nu}$ and $N_{\mathrm{eff}}$ which also allowed to significantly lower the cosmological neutrino mass bound. Due to today's precise CMB data of the Planck satellite \cite{Ade:2015xua} the degeneracy between $N_{\mathrm{eff}}$ and $\sum m_{\nu}$ is mainly eliminated such that a non-standard $N_{\mathrm{eff}}$ alone does not significantly weaken the neutrino mass bound anymore. The impact of a non-thermal neutrino distribution function on the neutrino mass bound however has to our knowledge not been studied after the first release of Planck data in 2013 \cite{Ade:2013zuv}.

This work is organized as follows: In section \ref{sec:Relic neutrino distribution} we present a parameterization for deviations from equilibrium of the neutrino distribution function. It describes out-of-equilibrium distributions that are not too far from the thermal
case. As examples, we apply our parameterization to non-standard distributions inspired by $\mu$- and $y$-type distortions of the cosmic photon spectrum \cite{Zeldovich:1969ff}. We present the impact of our parameterization on the CMB temperature anisotropy spectrum and the LSS matter power spectrum in section \ref{sec:Impact on the CMB and LSS}. By normalizing the distribution function  and adopting an effective neutrino mass, we show that our parameterization allows to weaken the neutrino mass bound by more than 100\%. We explicitly demonstrate this by performing a Markov chain Monte Carlo (MCMC) analysis in section \ref{sec:MCMC analysis} and finally conclude in section \ref{sec:Conclusions}.

\section{Relic neutrino distribution}
\label{sec:Relic neutrino distribution}

For massless neutrinos cosmological perturbation theory is entirely independent of the neutrino distribution function. All relevant quantities related to neutrinos -- e.g. the energy contrast or velocity divergence -- are already integrated over momentum. However, for massive neutrinos the Boltzmann hierarchy is momentum-dependent \cite{Ma:1995ey}. The equations for the zeroth and second multipole moments ($\Psi_{\nu,\ell}$) directly depend on the neutrino background-distribution function $f_{\nu}$, i.e.~(following the notation of \cite{Ma:1995ey})
\begin{equation}
\begin{aligned}
\dot{\Psi}_{\nu,0} & = - \frac{qk}{\epsilon} \Psi_{\nu,1} + \frac{1}{6} \dot{h} \frac{\mathrm{d}\ln f_{\nu}}{\mathrm{d}\ln q}, \\
\dot{\Psi}_{\nu,2} & = \frac{qk}{5 \epsilon}(2 \Psi_{\nu,1} - 3 \Psi_{\nu,3}) - \left( \frac{1}{15} \dot{h} + \frac{2}{5} \dot{\eta} \right) \frac{\mathrm{d} \ln f_{\nu}}{\mathrm{d}\ln q}.
\label{eq:BoltzmannHierarchy}
\end{aligned}
\end{equation}
Here, $q$ denotes comoving momentum and $\epsilon$ comoving neutrino energy, $k$  wave number, $h$ and $\eta$ are metric perturbations in the synchronous gauge, and overdots denote derivatives with respect to conformal time. The contributions to the energy-momentum tensor are derived by solving \eqref{eq:BoltzmannHierarchy} together with the $\ell=1$ and $\ell>2$ equations on a momentum grid, and integrating over momentum,~e.g.
\begin{align}
\delta \rho_{\nu} & = 4 \pi a^{-4} \int \mathrm{d}q \, q^2 \epsilon f_{\nu}(q) \Psi_{\nu,0}, \\
(\bar{\rho}_{\nu}+\bar{P}_{\nu})\sigma_{\nu} & = \frac{8 \pi}{3} a^{-4} \int \mathrm{d}q \, \frac{q^4}{\epsilon} f_{\nu}(q) \Psi_{\nu,2},
\label{integrated_quantities4}
\end{align}
for the energy density perturbation $\delta \rho_{\nu}$ and the anisotropic 
stress $\sigma_{\nu}$. $a$ denotes the scale factor. 
Hence, the neutrino distribution function $f_{\nu}$ has a direct impact on the evolution of massive neutrino perturbations which in turn have an impact on the photon and matter perturbations. Cosmological observables like the CMB anisotropy spectrum and the LSS matter power spectrum therefore depend on the neutrino distribution as well. 

Let us discuss how to parameterize non-standard neutrino distributions in a relatively model-independent way in the the following section.

\subsection{Parameterization of relic neutrino distribution}
\label{subsec:Parameterization of relic neutrino distribution}

According to the standard model of particle physics, neutrinos decouple from the cosmic plasma at around $T \sim$ 1 MeV. Since neutrinos are effectively massless at decoupling, they are usually assumed to follow an ultra-relativistic Fermi-Dirac (FD) distribution, which is even preserved as the neutrino momentum distribution at later times when neutrinos become non-relativistic,
\begin{equation}
f_{\nu}(x)= \frac{1}{e^{x}+1}.
\label{eq:Fermi-Dirac}
\end{equation} 

Here, we introduced the dimensionless comoving momentum $x=q/T_{\nu,0}$, where $T_{\nu,0}$ denotes the neutrino temperature today. Note that there are small distortions to the FD distribution \eqref{eq:Fermi-Dirac} due to $e^+-e^-$ annihilation which are however usually expressed in terms of an increased number of neutrino flavours, i.e. $N_{\text{eff}}=3.046$ \cite{Mangano:2005cc,deSalas:2016ztq}.

There are many exotic scenarios which could possibly alter the standard distribution function \eqref{eq:Fermi-Dirac} of neutrinos. Examples for this are: the decay or annihilation of some heavier particle into neutrinos (e.g. \cite{Kawasaki:1999na,Kawasaki:2000en,Cuoco:2005qr,GonzalezGarcia:2012yq}), non-standard neutrino interactions like the freeze-in of a massless Majoron (e.g. \cite{Basboll:2008fx,Oldengott:2014qra}) or decaying neutrinos (e.g. \cite{Starkman:1993ik,Hannestad:1999xy})\footnote{Non-standard interactions scenarios in general would however also change the free-streaming behavior of neutrinos \cite{Hannestad:2004qu} which would induce further modifications to the CMB and LSS that are not discussed here.}, or alternatively oscillations to sterile neutrino states \cite{Barbieri:1989ti,Foot:1995qk,Hannestad:2012ky,Hannestad:2013wwj,Hannestad:2015tea}. In order to keep the discussion as general as possible we do not want to focus on any specific scenario but aim to be as model-independent as possible. We therefore choose a parameterization very similar to \cite{Esposito:2000hi} and expand the distribution function in polynomials which are orthonormal with respect to the FD distribution, i.e. 
\begin{equation}
\int_0^\infty \mathrm{d}x \frac{1}{e^x+1}p_n(x) p_m(x)=\delta_{nm}.
\label{eq:orthonormal}
\end{equation}

This choice for the orthonormality relation is different from \cite{Cuoco:2005qr}, where the polynomials were chosen to be orthonormal to $x^2/(e^x+1)$. Since one of the aims of this work is to study the impact of the distribution function itself (i.e. of $f_{\nu}(x)$ and not of $x^2 f_{\nu}(x)$), we find our choice more appropriate for this purpose. The first six polynomials can be found in the appendix \ref{AppendixA}. 
The distribution function can then be written as
\begin{equation}
f_{\nu}(x)= \frac{1}{e^x+1} \sum_{n=0}^{\infty} c_n p_n(x),
\label{eq:expansion1}
\end{equation}
where the coefficients $c_n$ are given by
\begin{equation}
c_n= \int_0^{\infty} \mathrm{d}x f_{\nu}(x) p_n(x).
\label{eq:multipoles}
\end{equation}

Note that equation \eqref{eq:expansion1} is entirely model-independent since \textit{any} function can be decomposed into the orthogonal polynomials \eqref{eq:orthonormal}. In general however, a distribution function very different from the thermal FD distribution necessitates the inclusion of a high number of coefficients $c_n$ \eqref{eq:multipoles}. Nevertheless, by restricting our discussion to only a few moments in \eqref{eq:expansion1}, we can already describe a relatively large variety of distribution functions. Let us therefore consider only polynomials up to order $n=2$ in the following.

A disadvantage of the parameterization \eqref{eq:expansion1} is that for arbitrary choices of $c_0$, $c_1$ and $c_2$ the distribution can get negative values which is of course an unphysical feature. It is however easy to show that this unphysical behavior can be avoided by restricting the multipoles to $c_1 \leq 0.92 c_0$ and $c_2 \leq 1.06 c_0$ (see appendix \ref{AppendixA} for a  detailed discussion). In order to better control $c_1$ and $c_2$ it turns out to be more convenient to  factorize $c_0$ in \eqref{eq:expansion1} and interpret it as an overall normalization $N$. Without loss of generality, the expansion \eqref{eq:expansion1} up to $n=2$ can therefore be rewritten as
\begin{equation}
f_{\nu}(x)= N \cdot \frac{1}{e^x+1} \bigg ( p_0(x) + F_1 p_1(x) + F_2 p_2(x) \bigg ),
\label{eq:expansion2}
\end{equation} 
where $N \equiv c_0$ and $F_{1/2} \equiv (c_{1/2}/N)$. The conditions to have a positive distribution function therefore translate into
\begin{equation}
0 \leq F_1 < 0.92, \hspace{1cm} 0 \leq F_2 < 1.06.
\label{eq:conditions}
\end{equation}

\subsection{Examples}

As an example, we apply our expansion \eqref{eq:expansion2} to two types of familiar non-thermal features in distribution functions, namely $\mu$- and $y$-type distortions. Both of these distortions are known from the CMB spectrum, where they can appear due to inefficient ($y$-type) or efficient ($\mu$-type) Compton scattering with electrons \cite{Zeldovich:1969ff}. $\mu$-type distortions could furthermore be realized by neutrino chemical potentials, which would appear if the lepton asymmetry of our Universe was non-negligible, see e.g.~\cite{Simha:2008mt,Oldengott:2017tzj}. In that case, they should however not be interpreted as \textit{distortions} since a chemical potential is a thermal (but non-standard) feature of the neutrino distribution function. For $y$-type distortions there is however no obvious realization within the neutrino sector. In analogy to the Sunyaev-Zeldovich effect \cite{Zeldovich:1969ff}, such a distortion could in principle be produced by neutrinos scattering on some high-energetic particles after neutrino decoupling. However, without claiming any specific production mechanism of such a distortion, we consider it -- only for the purpose of testing how well the parameterization \eqref{eq:expansion2} can mimic non-standard distribution functions.

\begin{figure}
\centering
\begin{minipage}{0.49\textwidth}
\includegraphics[width=\textwidth]{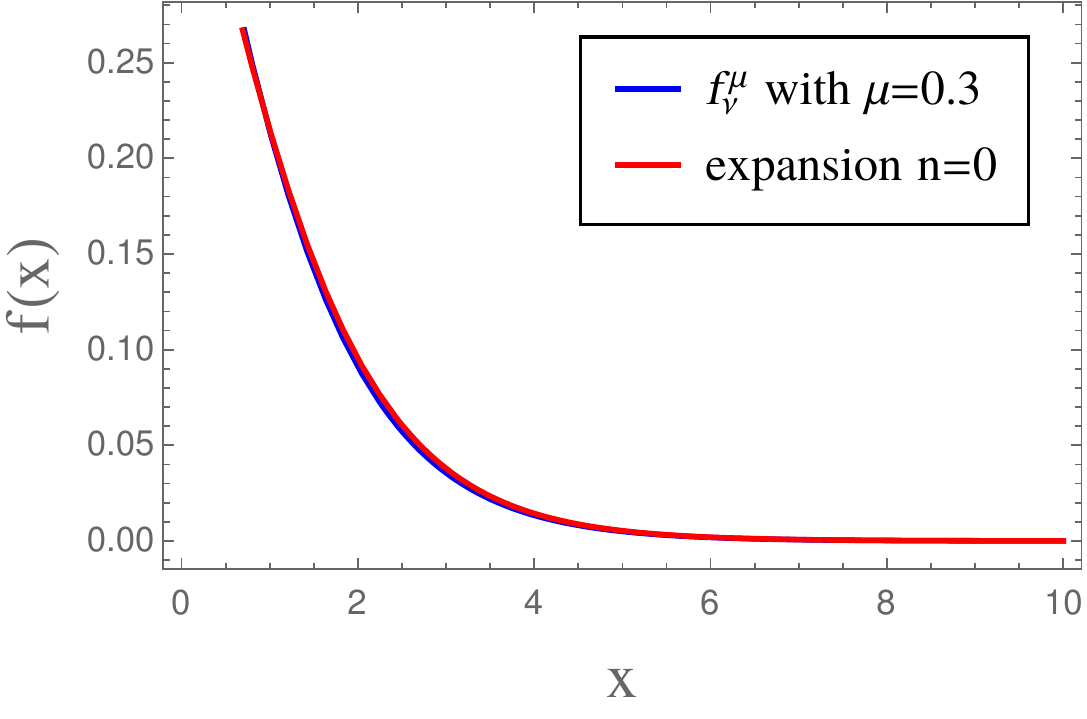}
\end{minipage}
\begin{minipage}{0.49\textwidth}
\includegraphics[width=\textwidth]{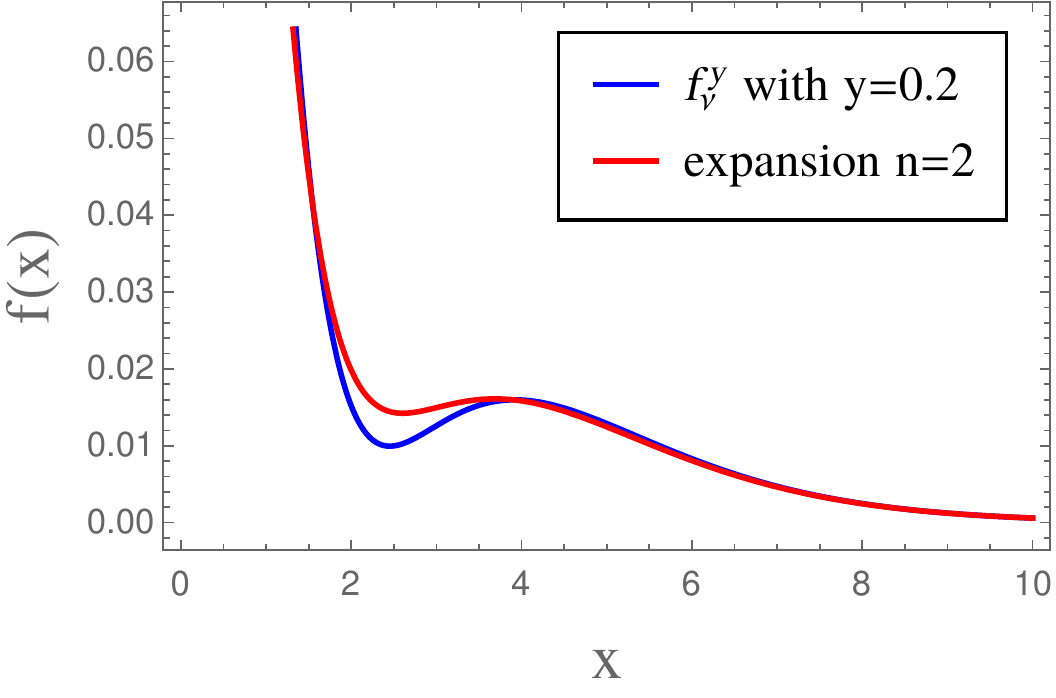}
\end{minipage}
\caption{\textit{Left:} Comparison of function \eqref{eq:FD+chem} for $\mu=0.3$ and the expansion \eqref{eq:expansion2} up to order $n=0$. \textit{Right:} Same, but with a $y$-type distortion \eqref{eq:ytype} and the expansion up to order $n=2$. }
\label{fig:ChemPot}
\end{figure}

Let us start with considering a $\mu$-type distortion,
\begin{equation}
f^{\mu}_{\nu}(x) = \frac{1}{e^{x+\mu}+1},
\label{eq:FD+chem}
\end{equation}  
where we exemplary chose $\mu=0.3$ here. In thermodynamic equilibrium $\mu$ denotes the ratio of chemical potential and temperature and after neutrino decoupling it becomes the 
so-called neutrino degeneracy parameter. 

The normalization $N$ in \eqref{eq:expansion2} can be calculated according to  \eqref{eq:multipoles} and is given by
\begin{equation}
N=c_0=0.665848.
\end{equation} 
The $F_1$ and $F_2$ parameters are much smaller than 1, which shows that they are very subdominant in comparison to the overall normalization $N$. Indeed, if we neglect $F_1$ and $F_2$ (i.e. we cut the expansion at $n >0$), we can see in the left panel of figure \ref{fig:ChemPot} that the distribution function resembles the exact expression in \eqref{eq:FD+chem} quite well. 

$y$-type distortions are described according to
\begin{equation}
f^{y}_{\nu}(x)= \frac{1}{e^x+1} \left[ 1+ y \frac{x e^x}{e^x+1} \left( x \frac{e^x-1}{e^x+1}-4 \right) \right].
\label{eq:ytype}
\end{equation}
Note that compared to $y$-type distortions of CMB photons there are some sign switches in \eqref{eq:ytype} due to the difference between bosons and fermions in the Kompaneets equation.

Exemplary we study the case of $y=0.2$ and find for the expansion parameters in \eqref{eq:expansion2}
\begin{equation}
N=c_0=0.499533, \hspace{1cm} F_1 = \frac{c_1}{N}= 7.921424 \cdot 10^{-8},\hspace{1cm} F_2 = \frac{c_2}{N}= 0.775381.
\label{eq:expansion_ytype}
\end{equation}

The right panel of figure \ref{fig:ChemPot} shows the exact expression \eqref{eq:ytype} with $y=0.2$ and the expansion \eqref{eq:expansion2} with the parameter values  \eqref{eq:expansion_ytype} in comparison. Even though the expansion does not accurately reproduce the exact expression near the local minimum, the overall consistency is reasonably good.

We conclude that our expansion can mimic $\mu$- and $y$-type distortions. However, as we include only quadratic polynomials in equation \eqref{eq:expansion2} our parameterization is of course restricted to certain functions and we cannot mimic every thinkable distribution function by equation \eqref{eq:expansion2}. Given a specific model for non-standard neutrino distributions it may be more useful to parameterize the distribution function in a different way.

This being said, we also want to point out another conclusion that we can draw from this section: Since the normalization $N$ has a very similar effect as the effective number of neutrino flavours $N_{\mathrm{eff}}$, we conclude that chemical potentials are degenerate with $N_{\mathrm{eff}}$. While this is a known fact for massless neutrinos (where the distribution function itself does not matter, as explained above), this has not been a priori clear for the case of massive neutrinos. Therefore, even with future experiments like CMB-S4 \cite{Abazajian:2016yjj} or Euclid \cite{Laureijs:2011gra,Amendola:2012ys} it will be impossible to disentangle the effect of neutrino chemical potentials and $N_{\mathrm{eff}}$. Note that this conclusion however only holds for the muon and tau neutrino, since the electron neutrino chemical potential has an additional impact on big bang nucleosynthesis (BBN) that changes the primordial helium abundance and has a much larger impact on the CMB anisotropy spectrum \cite{Oldengott:2017tzj}. 

\section{Impact of neutrino spectral distortions on the CMB and LSS}
\label{sec:Impact on the CMB and LSS}

Let us now study in detail the impact of the parameterization \eqref{eq:expansion2} of neutrino spectral distortions on the CMB temperature anisotropy spectrum and the LSS matter power spectrum. In order to do so we should make a short detour to BBN.

The impact of neutrinos on BBN is two-fold: The electron-neutrino distribution function directly impacts BBN as it determines the neutron-to-proton ratio at the onset of BBN at $T=\mathcal{O}$(0.1 MeV). The impact of the muon- and tau-neutrinos is more indirect as they only impact BBN through the expansion rate, which depends on the energy density of the ultra-relativistic neutrinos and 
hence can be parametrized by $N_{\text{eff}}$. Hence, in principle a non-standard neutrino distribution function changes the BBN prediction for the abundance of primordial helium, which in turn has an impact on the CMB anisotropy spectrum. However, in most thinkable scenarios the neutrino distribution function would probably still be FD at BBN: Before neutrino decoupling at $T=\mathcal{O}$(1 MeV) any non-thermal feature in the neutrino distribution function would be washed out quickly by weak interactions \footnote{Note that this does not hold for chemical potentials -- discussed in the previous section -- which are a thermal (but non-standard) feature in the neutrino distribution function.}. The window for non-standard physics to alter the neutrino distribution \textit{before the onset of BBN} is therefore restricted to the cosmic temperature range $T=\mathcal{O}$(0.1 MeV) to  $T=\mathcal{O}$(5 MeV). For a detailed analysis of such a scenario see \cite{Kawasaki:1999na,Kawasaki:2000en}. However, in the general framework of this work it seems to be reasonable to restrict our discussion to scenarios where the neutrino distribution only becomes non-standard \textit{after} the onset of BBN. Some scenarios could nevertheless still modify BBN through an enhanced $N_{\text{eff}}$, e.g. the decay of a particle of keV mass.  
Note that a particle of $> \mathcal{O}$(MeV) mass would in contrast only enhance $N_{\mathrm{eff}}$ after BBN, as it would already be non-relativistic and therefore not enhance the relativistic energy density at BBN. As we discuss in the next section, we focus on scenarios that do not alter $N_{\mathrm{eff}}$ but only the way how neutrinos are distributed. 

\subsection{Degeneracy of $N_{\text{eff}}$ and spectral distortions --- normalization of the distribution function}

It is already well-known that the neutrino distribution function is degenerate with $N_{\mathrm{eff}}$, see e.g.~\cite{Cuoco:2005qr}. In order to study the required sensitivity of current or future experiments to disentangle the effect of $N_{\mathrm{eff}}$ and the distribution function, we are interested in finding out if there is a \textit{unique} feature of the neutrino distribution function in the CMB anisotropy spectrum and the LSS matter power spectrum. In order to switch off the effect due to $N_{\mathrm{eff}}$ it is  useful to choose the normalization $N$ \eqref{eq:expansion2} such that $N_{\text{eff}}$ always gives the same value. This can be realized if 
\begin{equation}
N \cdot \int \mathrm{d}x \frac{x^3}{e^{x}+1}  \bigg ( p_0(x) + F_1 p_1(x) + F_2 p_2(x) \bigg ) \overset{!}{=} \int \mathrm{d}x \frac{x^3}{e^{x}+1},
\label{eq:Normalization}
\end{equation}
which is fulfilled by the following choice of $N$ 
\begin{equation}
N= \frac{5.682197}{6.825012 + 18.239415 F_1 + 17.420046 F_2}.
\label{eq:Normalization}
\end{equation}

\begin{figure}
\centering
\begin{minipage}{0.49\textwidth}
\includegraphics[width=\textwidth]{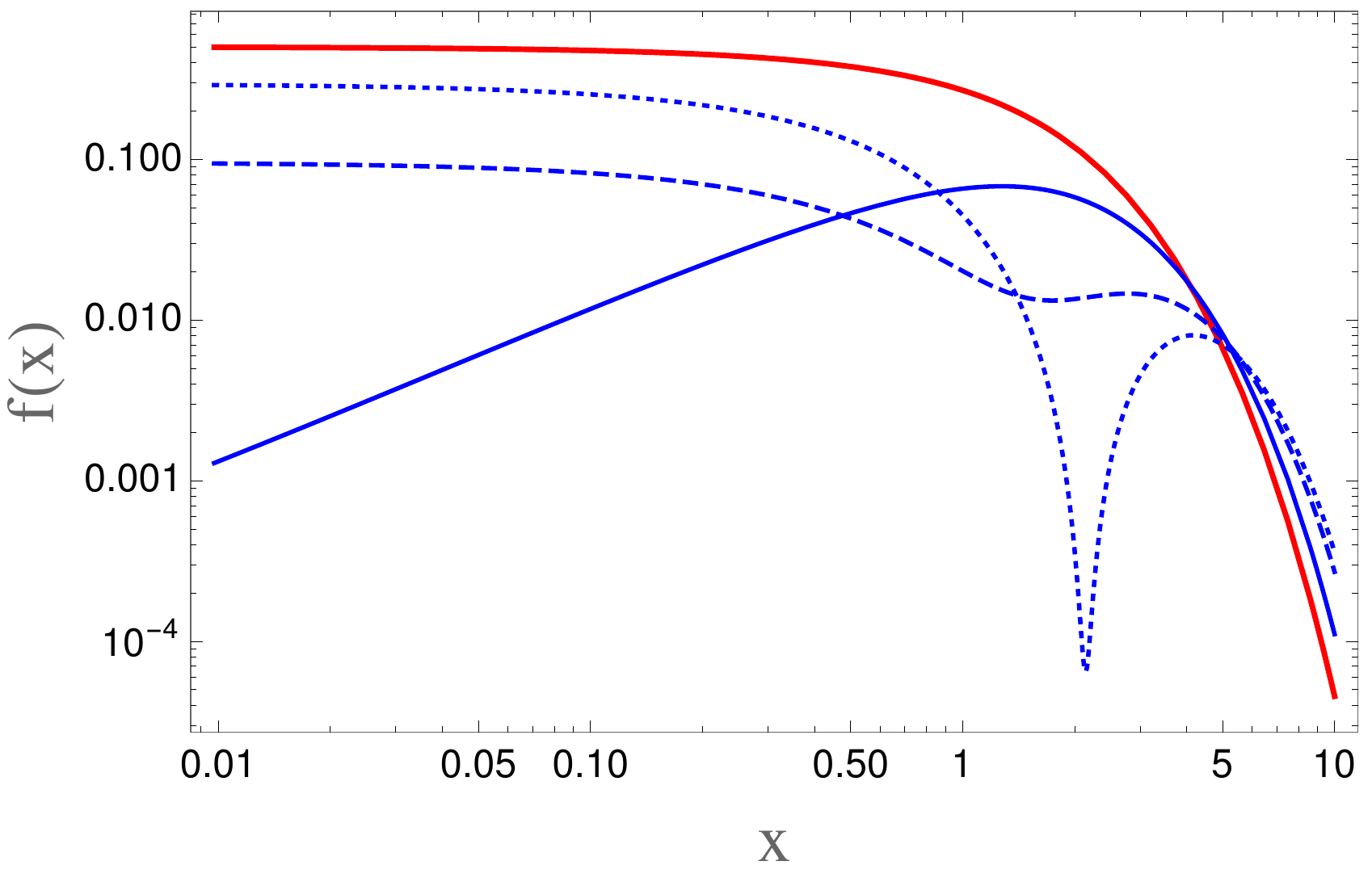}
\end{minipage}
\begin{minipage}{0.49\textwidth}
\includegraphics[width=\textwidth]{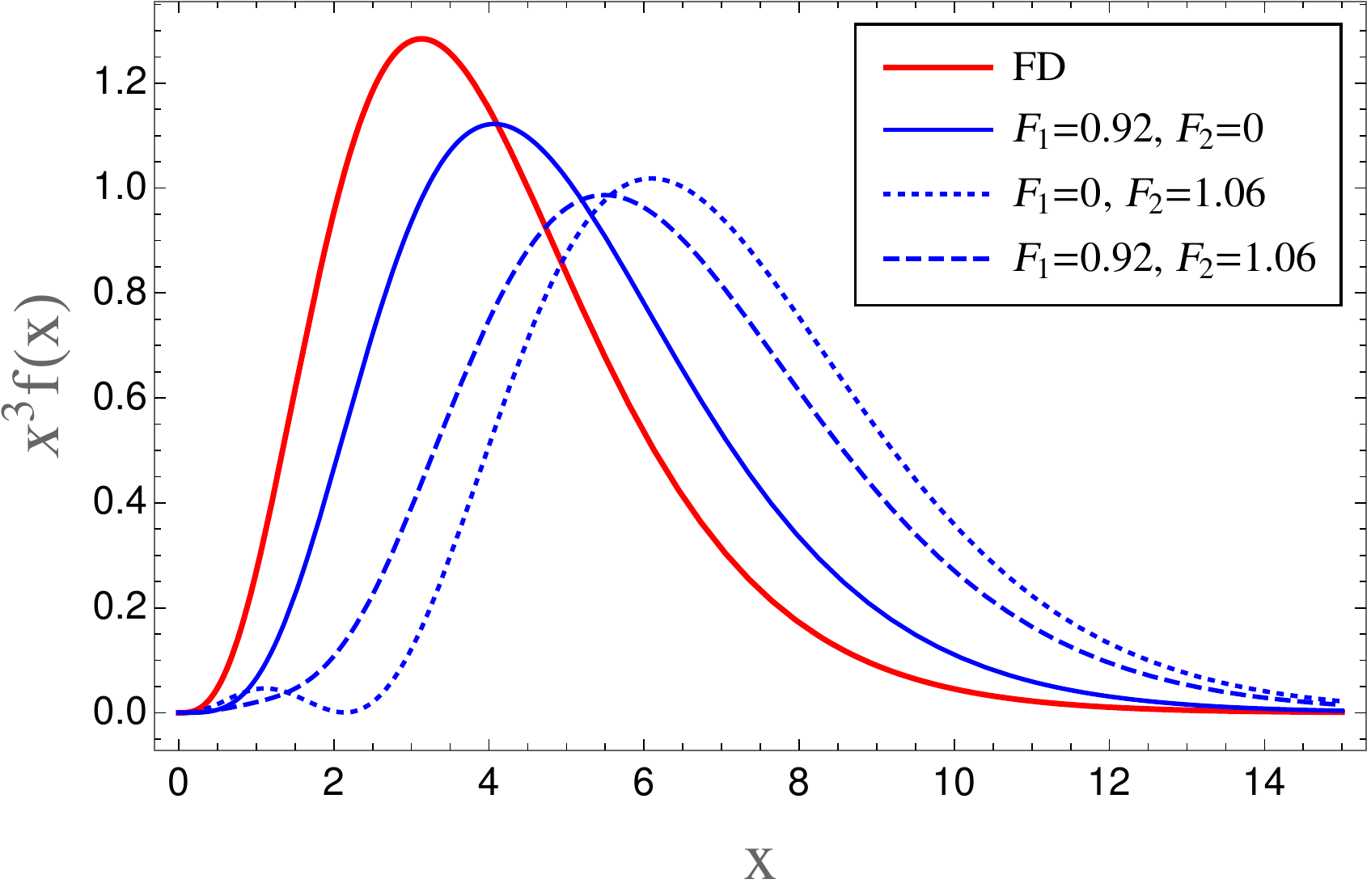}
\end{minipage}
\caption{Comparison of different parameter combinations for $F_1$ and $F_2$ \eqref{eq:expansion2} for the normalized neutrino distribution function $f_{\nu}(y)$ (left) and $y^3 f_{\nu}(y)$ (right).}
\label{fig:distribution}
\end{figure}

\begin{figure}
\centering
\begin{subfigure}{\textwidth}
\includegraphics[width=\textwidth]{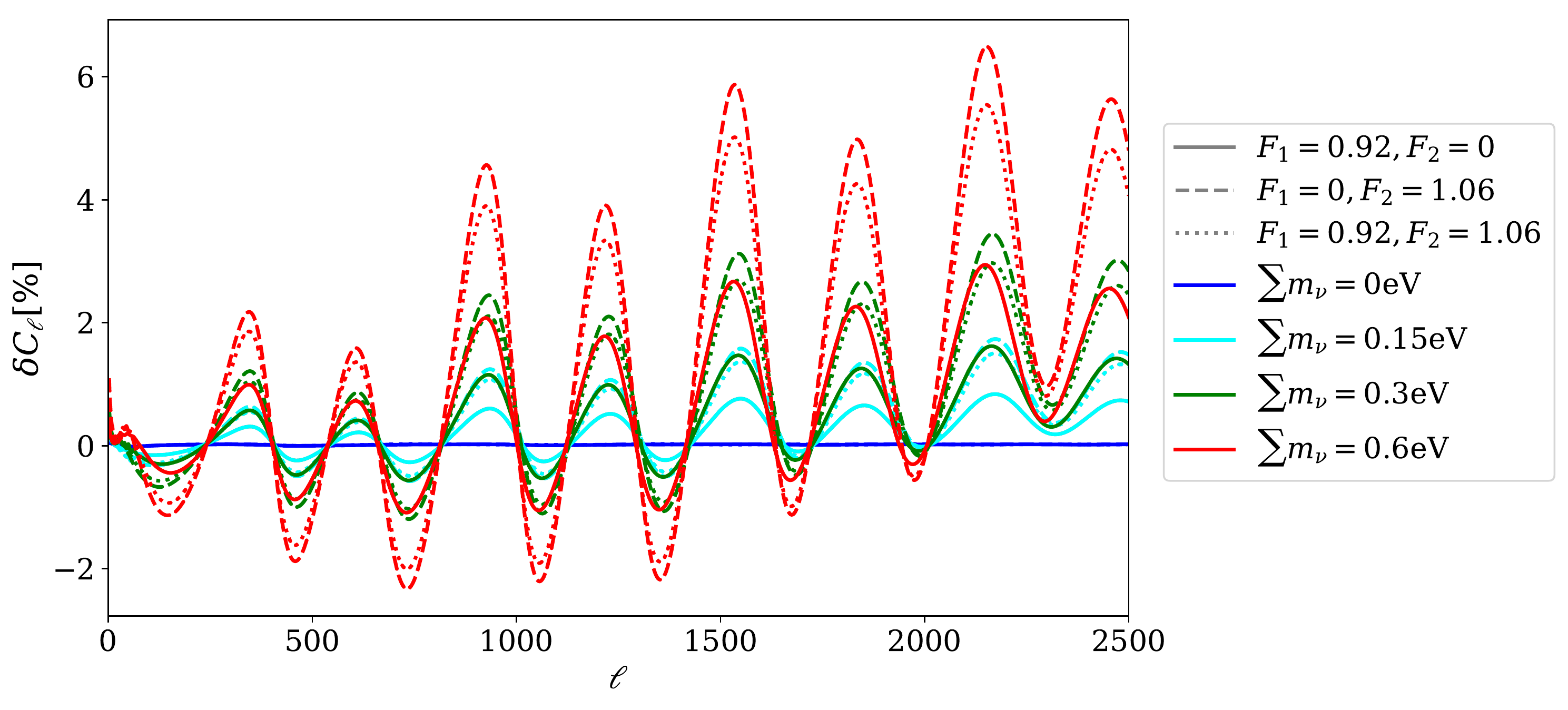}
\end{subfigure}
\begin{subfigure}{\textwidth}
\includegraphics[width=\textwidth]{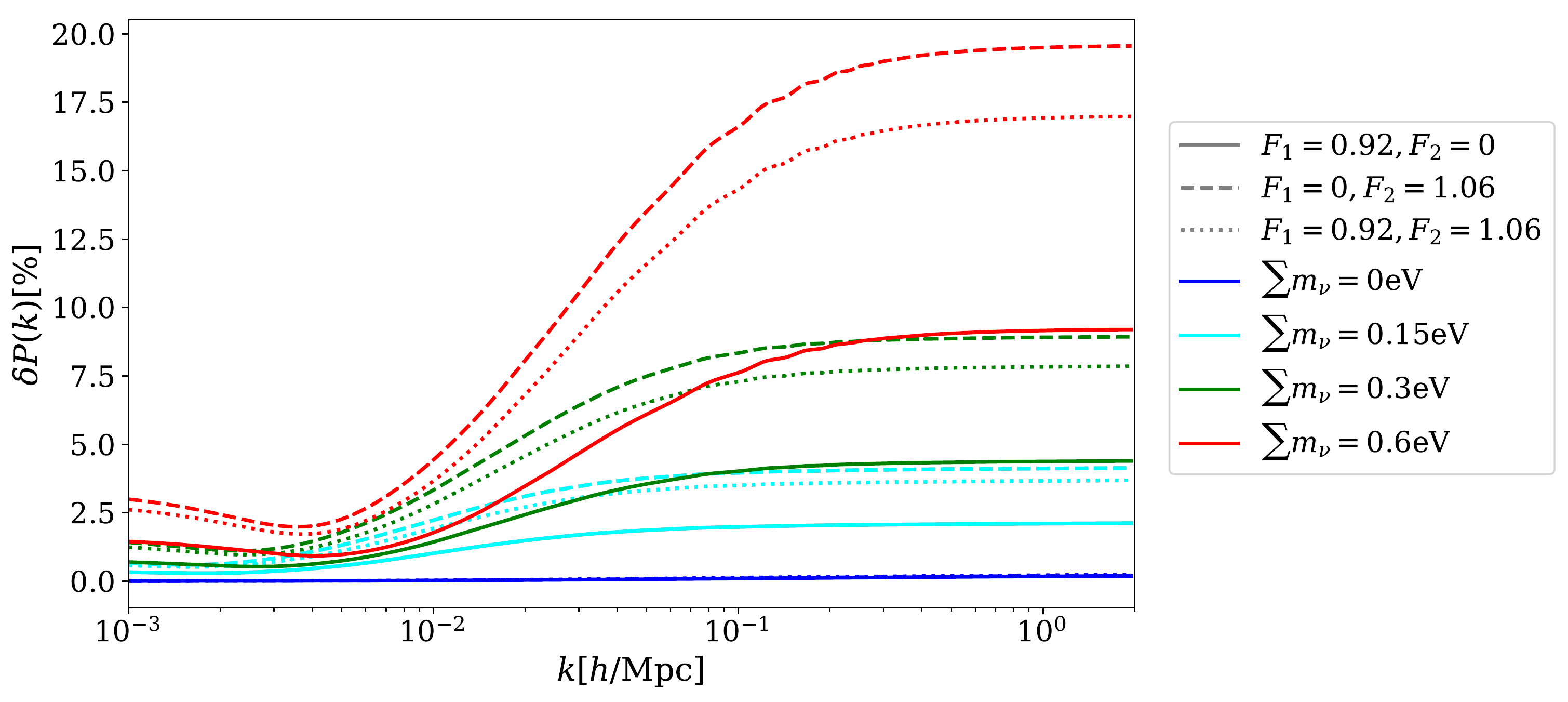}
\end{subfigure}
\caption{Relative difference in the CMB temperature anisotropy spectrum (top) and LSS matter power spectrum (bottom) between a standard FD neutrino distribution and the parameterization \eqref{eq:expansion2} with the normalization \eqref{eq:Normalization} for different neutrino masses and choices of $F_1$ and $F_2$.}
\label{fig:delta_ClsPk1}
\end{figure}

Such a scenario (i.e.\ a non-thermal distribution function but an unaltered $N_{\mathrm{eff}}$) could be realized for example by the freeze-in of a massless Majoron \cite{Basboll:2008fx,Oldengott:2014qra}, decaying neutrinos \cite{Starkman:1993ik,Hannestad:1999xy} or oscillations to sterile neutrinos \cite{Barbieri:1989ti,Foot:1995qk,Hannestad:2013wwj}. 

In figure \ref{fig:distribution} we show the normalized \eqref{eq:Normalization} neutrino distribution function \eqref{eq:expansion2} as well as $x^3 f_{\nu}(x)$ for different choices of $F_1$ and $F_2$.
All shown cases of spectral distortions give rise to an increase of the mean neutrino momentum as compared to the FD spectrum. We can see that the dotted line with $\lbrace F_1=0,F_2=1.06 \rbrace$ is the most efficient parameter combination in shifting the average momentum.

\subsection{Implementation and cosmological observables}

We implemented the parameterization \eqref{eq:expansion2} with the normalization given in \eqref{eq:Normalization} into the linear Einstein-Boltzmann solver {\sc class} \cite{Blas:2011rf}. Note that for parameter combinations with $F_1 \neq 0$ and $F_2 \neq 0$ \textit{at the same time} the precision of the numerical momentum-integration method has to be slightly increased in order to obtain precise results. Here we focus our study on mass degenerate neutrinos, i.e. $m_1=m_2=m_3$, as we are only interested in the upper limit on the sum of all neutrino masses. The base model is a spatially flat, $\Lambda$ cold dark matter model with scale-free 
primordial curvature fluctuations and without primordial gravitational waves, i.e.\ todays vanilla model. 

Figure \ref{fig:delta_ClsPk1} shows the relative percent difference in the angular power 
spectrum of CMB temperature anisotropy ($C_\ell$) and the LSS matter power spectrum ($P(k)$)
between a non-standard neutrino distribution function \eqref{eq:expansion2} and the standard FD distribution, i.e. 
\begin{equation}
\delta C_{\ell}= \frac{C_{\ell}-C^{FD}_{\ell}}{C^{FD}_{\ell}}
\end{equation}
and equivalently for the matter power spectrum $P(k)$. The dark blue curve ($\sum m_{\nu}=0$ eV) is only plotted as a numerical test to confirm that for massless neutrinos cosmological observations are indeed independent of the distribution function. 

As expected, the impact of a non-thermal distribution function is increasing for increasing neutrino masses (in fact approximately proportional to the sum of neutrino masses). 
Remarkably the relative difference in the LSS matter power spectrum is up to 20\% for the parameter combinations under consideration, whereas it is about a factor of 3 smaller -- but still significant -- in the CMB anisotropy spectrum. Besides a modulation of the angular power spectrum, both in the CMB and in the LSS we can observe an 
enhancement of structure at small length scales. This shows that the neutrino distribution function has indeed an impact on the the CMB and LSS which is not degenerate with $N_{\mathrm{eff}}$. The remaining question is now whether the effects 
shown in figure \ref{fig:delta_ClsPk1} are \textit{unique} features of the neutrino distribution function or 
if they can be mimicked by other cosmological parameters.

\subsection{Determining an effective neutrino mass}

Of course, the normalization $N$ \eqref{eq:Normalization} that we chose in the last section fixes the neutrino energy density at early times when neutrinos are relativistic. There is no way to find a normalization such that the energy density of the non-thermal distribution \eqref{eq:expansion2} is \textit{always} the same as in the standard FD case. 
However, the energy density at late times is not only determined by the distribution function but also by the neutrino mass. The CMB anisotropy spectrum of figure \ref{fig:delta_ClsPk1} also shows that the impact of a given neutrino mass can be mimicked to some extent by a different neutrino mass \textit{and} another corresponding distribution function: For example, the red solid line ($\sum m_{\nu}=0.6$ eV, $F_1=0.92$, $F_2=0$) looks very similar to the green dotted line ($\sum m_{\nu}=0.3$ eV, $F_1=0.92$, $F_2=1.06$). This motivates the question if we can eliminate the signal of a non-thermal neutrino distribution in figure \ref{fig:delta_ClsPk1} by adapting the neutrino mass accordingly. In other words, we want to determine the neutrino mass of the non-thermal distribution $m^*_{\nu}$, such that 
today's $\rho_\nu (m^*_\nu) = \rho^{FD}_\nu (m_\nu)$. Then the following equation is fulfilled
\begin{equation}
\int_0^\infty \mathrm{d}x  \, x^2 \sqrt{x^2+\frac{m_{\nu}^2}{T_{\nu 0}^2}} \, \left( \frac{1}{e^x+1} \right) \overset{!}{=} N \int_0^\infty \mathrm{d}x \, x^2 \sqrt{x^2+\frac{{m_{\nu}^*}^2}{T_{\nu 0}^2}} \, \frac{1}{e^x+1} \bigg ( p_0(x) + F_1 p_1(x) + F_2 p_2(x) \bigg ),
\label{eq:meff}
\end{equation}
where $N$ is given in equation \eqref{eq:Normalization}. 

\begin{figure}
\centering
\begin{subfigure}{\textwidth}
\includegraphics[width=\textwidth]{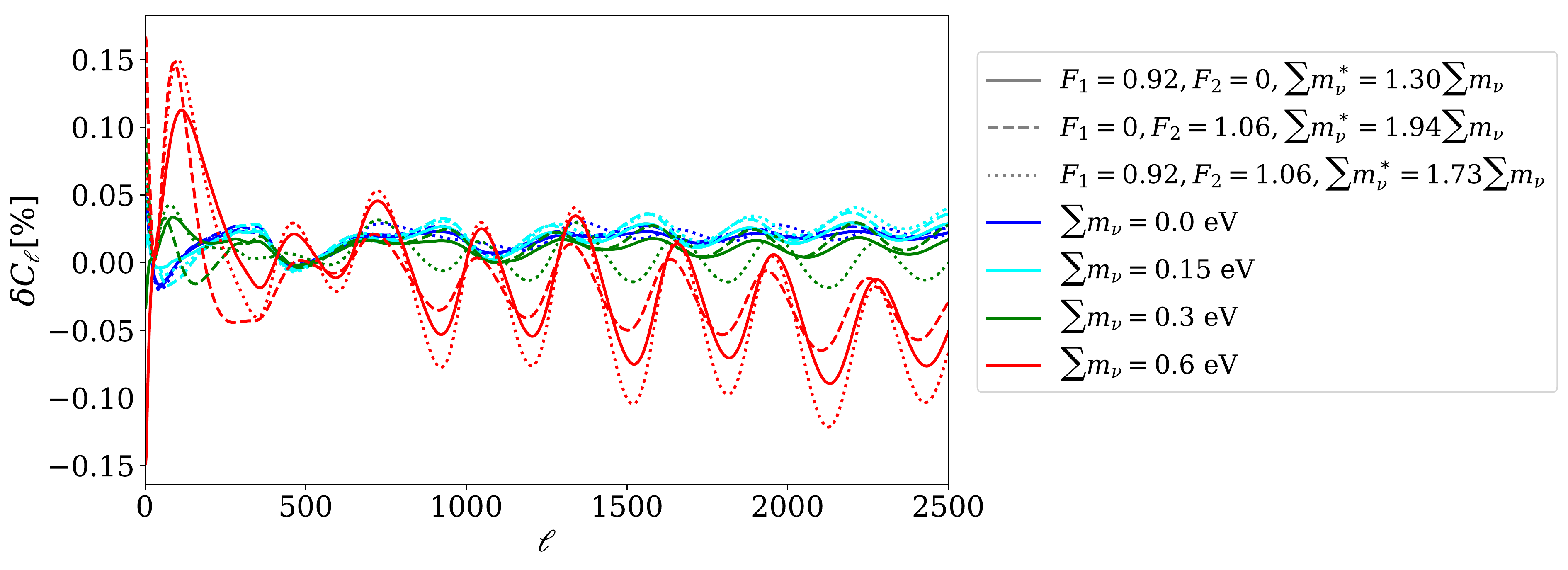}
\end{subfigure}
\begin{subfigure}{\textwidth}
\includegraphics[width=\textwidth]{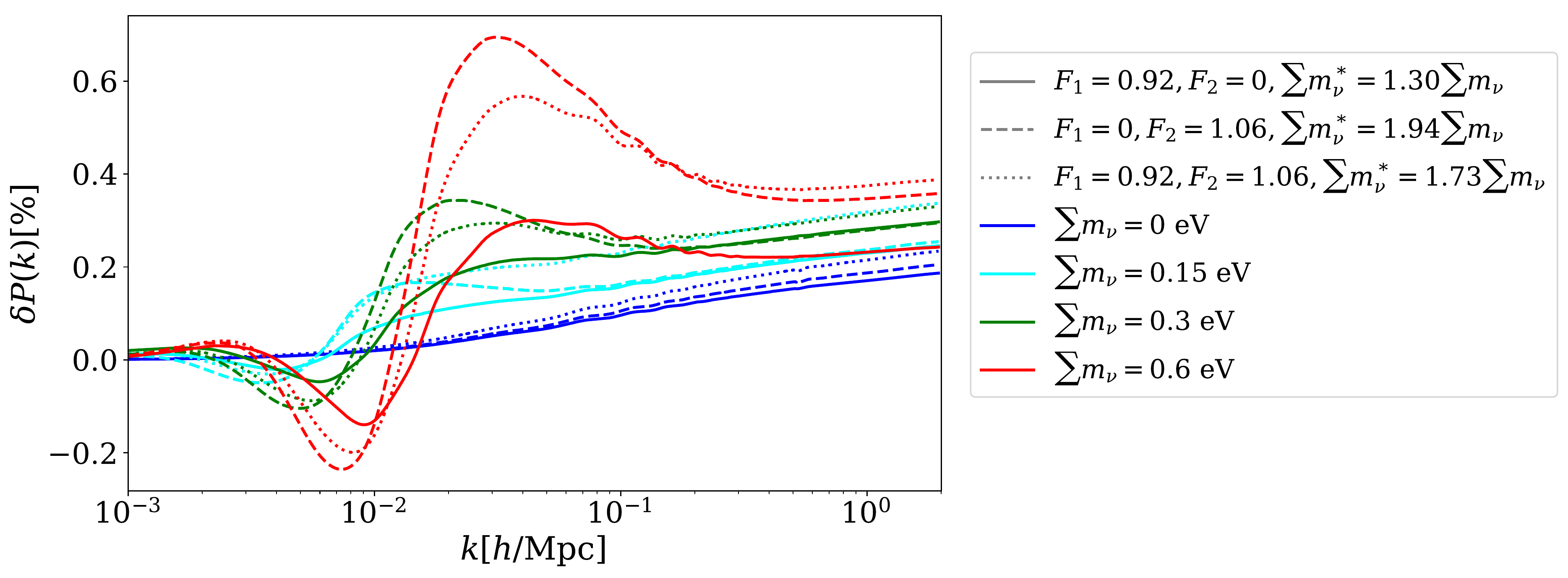}
\end{subfigure}
\caption{Same as figure \ref{fig:delta_ClsPk1}, but with the neutrino mass of the non-thermal distribution adapted according to \eqref{eq:meff2}.}
\label{fig:delta_ClsPk1_meff}
\end{figure}

By approximately solving equation \eqref{eq:meff} we find the following relations for the parameter combinations used in figure \ref{fig:delta_ClsPk1}:
\begin{equation}
\begin{aligned}
F_1 &= 0.92, &F_2 &= 0 &: \hspace{1cm} & m_{\nu}^* = 1.30 \cdot m_{\nu},  \\
F_1 &= 0, &F_2 &= 1.06 &: \hspace{1cm} & m_{\nu}^* = 1.94 \cdot m_{\nu}, \\
F_1 &= 0.92, &F_2 &= 1.06 &:\hspace{1cm} & m_{\nu}^* = 1.73 \cdot  m_{\nu}.
\end{aligned}
\label{eq:meff2}
\end{equation}

Figure \ref{fig:delta_ClsPk1_meff} shows how the signal presented in figure \ref{fig:delta_ClsPk1} changes when adapting the neutrino masses according to \eqref{eq:meff2}. After adapting the neutrino masses the remaining effect of the spectral distortions drops below the 1\% level in both,
temperature anisotropy spectrum and LSS matter power spectrum. Note that the numerical precision of our calculations are at the level of $0.02 \%$ for the CMB anisotropy spectrum and about $0.2 \%$ for the LSS matter power spectrum (dark blue curves). In order to track down how much of the signal in figure \ref{fig:delta_ClsPk1_meff} is due to the real unique feature of the neutrino distribution function and how much is simply numerical noise, we would have to tune the numerical precision of {\sc class}. However, the remaining signal in fig. \ref{fig:delta_ClsPk1_meff} is anyhow below the precision of current experiments and its detection therefore currently unfeasible. 

We have just shown that the neutrino distribution function is not only degenerate with $N_{\mathrm{eff}}$ but also with the neutrino mass. This has interesting consequences on the robustness of cosmological neutrino mass bounds: By abandoning the assumption that relic neutrinos follow a thermal FD distribution precisely, we can weaken cosmological  neutrino mass bounds. As we have shown, cosmological observations almost exclusively care about the time when most of the neutrinos become non-relativistic. This time is not only determined by the neutrino mass but also by the neutrino distribution function. Even when keeping the neutrino energy density at early times fixed (i.e. when $N_{\mathrm{eff}}$ is fixed), we can still enhance the neutrino average momentum such that neutrinos are allowed to be much heavier than in the standard framework. For the parameterization chosen in this work \eqref{eq:expansion2} we expect the neutrino mass bounds to become weaker by more than $90 \%$ (for $\lbrace F_1=0,F_2=1.06 \rbrace$, as shown in figure \ref{fig:delta_ClsPk1_meff}). 
We however want to strongly emphasize here that the amount by which the neutrino mass bound is changed clearly depends on the specific model under consideration. Some models with a non-thermal neutrino distributions may lead to a weaker effect, while other models may even relax the neutrino mass bound by much more than the $90 \%$ found here.

\section{MCMC analysis}
\label{sec:MCMC analysis}

In order to demonstrate on real data that we can lower the neutrino mass bound by allowing a non-thermal neutrino distribution, we perform an MCMC analysis with the MCMC engine Montepython \cite{Audren:2012wb,Brinckmann:2018cvx}. 
We study the two following data sets representing conservative and more stringent cosmological 
mass bounds:

\begin{description}
\item[TT+lowP] Temperature anisotropy spectrum and low-$\ell$ polarisation from the Planck 2015 data \cite{Ade:2015xua}.
\item[TT+lowP+BAO] Same as above plus measurements of the baryon acoustic oscillation peak scale as measured by 6DF~\cite{Beutler:2011hx}, BOSS LOWZ and CMASS~\cite{Anderson:2013zyy,Cuesta:2015mqa}, and the SDSS Main Galaxy Sample~\cite{Ross:2014qpa}.
\end{description}

As in the previous section, we study the case of mass-degenerate neutrinos since the neutrino mass-splittings are still close enough to be negligible with current data. We apply flat priors on all cosmological parameters and adopt the Gelman-Rubin convergence criterion $R<0.01$.

\begin{figure}
\centering
\includegraphics[width=0.9\textwidth]{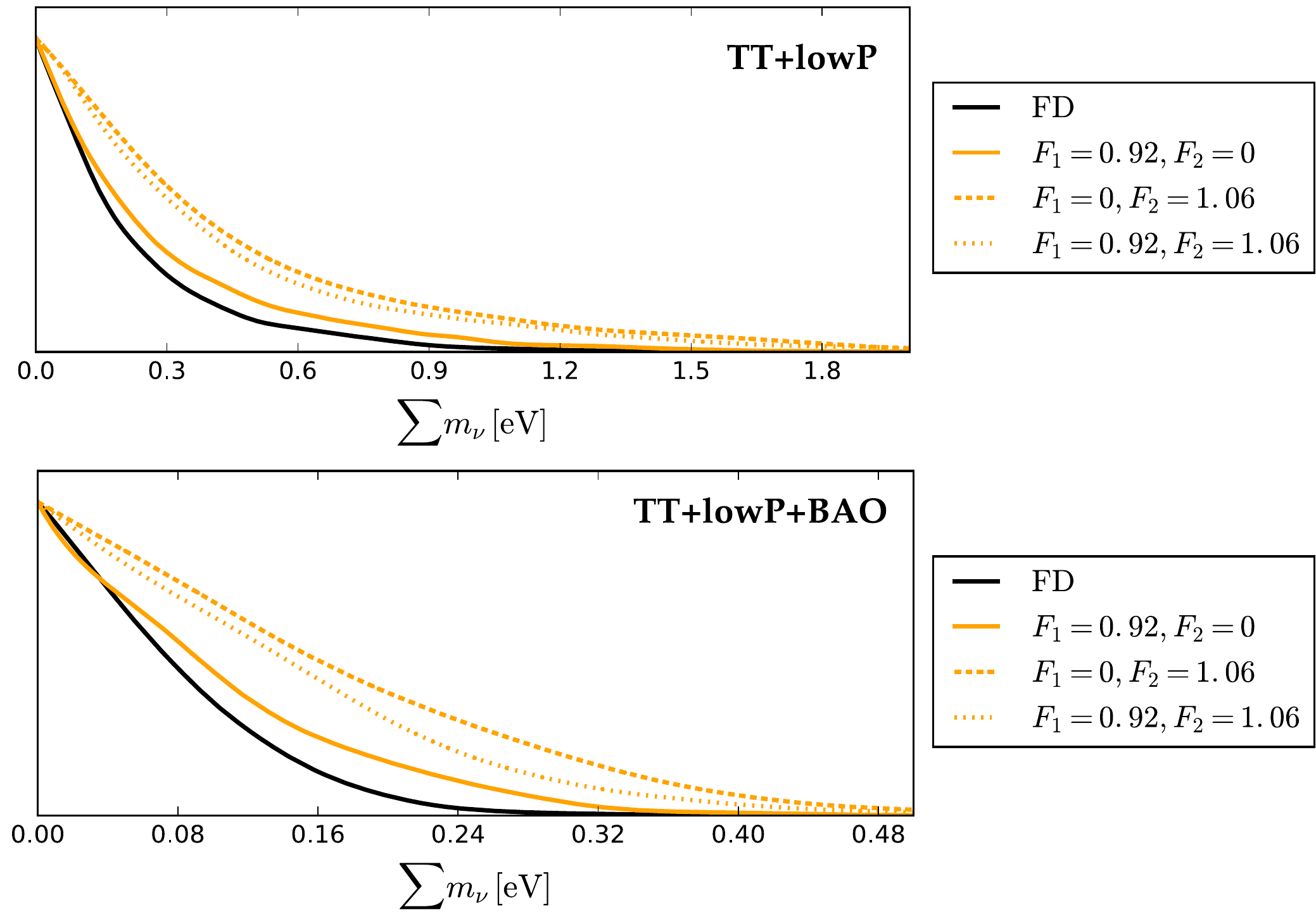}
\caption{Posterior distribution of the neutrino mass sum $\sum m_{\nu}$ for different choices for different choices of the multipoles $\lbrace 
F_1,F2\rbrace$ \eqref{eq:expansion2} and the normalization $N$ fixed to \eqref{eq:Normalization}, in comparison to the standard FD distribution. In all cases $N_{\text{eff}} = 3.046$.}
\label{fig:massbunds}
\end{figure}

We study different distribution functions, where we choose the same combinations of the expansion parameters $\lbrace F_1,F_2 \rbrace$ \eqref{eq:expansion2} as in figure  \ref{fig:delta_ClsPk1_meff} and equation \eqref{eq:meff2}. The overall normalization $N$ is thereby again fixed according to equation \eqref{eq:Normalization}. This also guarantees that we do not introduce any inconsistency with the CMB constraints on $N_{\text{eff}}$, which is fixed to $N_{\text{eff}}= 3.046$ in this analysis.
Note that a global fit, where $F_1$ and $F_2$ and $m_{\nu}$ are treated as free parameters, does not provide any further insight: Since at $m_{\nu}=0$ cosmic perturbation theory is blind to the relic neutrino distribution, $F_1$ and $F_2$ are entirely unconstrained in such an analysis. This also implies that future cosmological data will only have constraining power on the relic distribution function if they turn out to exclude $\sum m_{\nu}=0$ or if the neutrino masses eventually will be measured by laboratory experiments.

 As already noted in section \ref{sec:Impact on the CMB and LSS}, the precision of the numerical integration method for massive neutrinos in {\sc class} needs to be enhanced in order to properly perform momentum integrations. As a comparison we also performed an MCMC run assuming the standard FD distribution. Our results for the mass constraints in this run agree well with the ones by the Planck collaboration \cite{Ade:2015xua}, which are however weaker than the ones quoted in the introduction, since the Planck 2018 data \cite{Aghanim:2018eyx} are not publicly available at the time of finishing this work.

Figure \ref{fig:massbunds} shows the posterior distribution of the neutrino mass sum $\sum m_{\nu}$ for the different parameter combinations. 
We clearly see how the posterior stretches, depending on the assumptions on the neutrino 
momentum distribution. The 95\% CL upper bounds for the different parameter combinations and different data sets are reported in table \ref{tab:massbounds}. The mass bounds we find in our analysis confirm the prediction of equation \eqref{eq:meff2}, i.e. depending on the choices of the parameters $\lbrace F_1, F_2 \rbrace$ we can lower the bound by more than 90\%, 
in fact by even more than a factor of two in the case of the more stringent bounds (0.37 eV instead of 0.18 eV).

\begin{table}
\centering
\def\arraystretch{1.5}
\begin{tabular}{|c||c|c|}
\hline 
 & TT+lowP (95 \% CL) & TT+lowP+BAO (95 \% CL) \\ 
\hline 
\hline
 FD  & $\sum m_{\nu} < 0.73$ eV & $\sum m_{\nu} < 0.18$ eV \\
\hline 
$F_1=0.92,F_2=0$ & $\sum m_{\nu} < 0.95$ eV & $\sum m_{\nu} < 0.26$ eV \\ 
\hline 
$F_1=0,F_2=1.06$ & $\sum m_{\nu} < 1.45$ eV & $\sum m_{\nu} < 0.37$ eV \\ 
\hline 
$F_1=0.92,F_2=1.06$ & $\sum m_{\nu} < 1.34 $ eV & $\sum m_{\nu} < 0.32$ eV \\ 
\hline 
\end{tabular} 
\caption{95\% CL upper bounds on the sum of neutrino masses for different choices of the multipoles $\lbrace 
F_1,F_2\rbrace$ \eqref{eq:expansion2} and the normalization $N$ fixed to \eqref{eq:Normalization}, in comparison to the standard FD distribution. All upper limits are based on Planck 2015 likelihoods \cite{Ade:2015xua}.}
\label{tab:massbounds}
\end{table}

\section{Conclusions}
\label{sec:Conclusions}

In this work, we studied the impact of a non-thermal relic neutrino distribution on cosmological observables, i.e. the CMB temperature anisotropy spectrum and the LSS matter power spectrum. 
Any process that increases the mean momentum of cosmic relic neutrinos could give 
rise to an increased cosmological upper bound on the sum of the neutrino masses. 
Two types of processes can be imagined. Either a process that modifies the bulk of cosmic relic 
neutrinos, or a rare effect that affects only a small number of cosmic neutrinos but injects a lot of 
energy, e.g. a post-BBN decay of a heavy particle into high-energy neutrinos.
Our formalism is well suited to study the former type of effects, but the latter one 
requires much more care with the numerical accuracy of the Boltzmann codes to account for 
the hierarchy in neutrino energy scales. Here we have focused on the former type of effects.  

In order to parameterize non-thermal distributions in a model-independent way, we expanded the distribution function in orthonormal polynomials where we cut the expansion at order $n \geq 3$ \eqref{eq:expansion2} for simplicity. This introduces three parameters, $N$, $F_1$ and $F_2$. We demonstrated that our parameterization can mimic $\mu$-distortions and $y$-inspired distortions very well. In order to figure out if the distribution function has a \textit{unique} imprint in the CMB and the LSS we chose the normalization $N$ such that it always reproduces the standard value of $N_{\mathrm{eff}}$. A realization of such a scenario would be any (non-standard) physics that redistribute neutrinos without injecting a too large amount of energy into the neutrino sector. Examples of this are e.g. non-standard neutrino interactions like the freeze-in of a massless Majoron (e.g. \cite{Basboll:2008fx,Oldengott:2014qra}), decaying neutrinos (e.g. \cite{Starkman:1993ik,Hannestad:1999xy}), or oscillations of active to sterile neutrinos \cite{Barbieri:1989ti,Foot:1995qk,Hannestad:2012ky,Hannestad:2013wwj,Hannestad:2015tea}. After normalizing the distribution function, the remaining effect is up to 6\% in the CMB temperature anisotropy spectrum, and up to 20\% in the LSS matter power spectrum, depending on the choices of $\lbrace F_1, F_2 \rbrace$ and $\sum m_{\nu}$. For different choices of the parameters $\lbrace F_1, F_2\rbrace$ we furthermore calculated an effective neutrino mass $m_{\nu}^*$ that allows to also have the same energy density \textit{at late times} as the standard FD distribution with mass $m_{\nu}$, see equations  \eqref{eq:meff} and \eqref{eq:meff2}. When adjusting the neutrino mass of the non-thermal distribution to the effective mass $m_{\nu}^*$ the signal in the CMB and LSS shrinks down below the percent level. This means that the class of spectral distortions studied in this work has almost no unique observable imprint, since it is almost entirely degenerate with $N_{\mathrm{eff}}$ and $\sum m_{\nu}$. While the degeneracy to $N_{\mathrm{eff}}$ has already been well-known, the intrinsic degeneracy to $\sum m_{\nu}$ has been much less communicated and --- in our opinion --- overseen in discussions about the robustness of cosmological neutrino mass bounds. By performing an MCMC analysis we have demonstrated that the cosmological neutrino mass bound can be relaxed by more than 90\% \textit{for the parameterization studied in this work}. Note however that our parameterization \eqref{eq:expansion2} is constructed in such a way that it only describes distribution functions that are still relatively close to the standard FD distribution. Adding higher polynomials to the parameterization could further relax the mass bounds and assuming a drastically different distribution function could in principle relax the mass bounds in a drastic way. In general, given a specific model it has to be tested from case to case if our parameterization \eqref{eq:expansion2} is sufficient in mimicking the neutrino distribution and by which amount the neutrino mass bound would be relaxed. 

To sum up, we showed how much the cosmological neutrino mass bound relies on our assumption about the relic neutrino distribution function. This stands in contrast to claims that the mass bound is entirely robust. %We however do not intend to claim the opposite, i.e. that the cosmological neutrino mass bound would not be robust. 
Indeed, assuming that neutrinos follow a thermal FD distribution is --- on theoretical grounds --- reasonable. But from an observationally agnostic point of of view, we have never measured the relic neutrino momentum distribution (and will not in the near future), so we should  keep in mind that the cosmological mass bound and the constraints on $N_{\text{eff}}$ strongly depend 
on this assumption. Direct mass detection experiments like KATRIN \cite{Osipowicz:2001sq} are therefore still of great importance, even though their sensitivity to the neutrino mass is already beaten by cosmological observations \textit{under the assumption of thermally distributed relic neutrinos}. If KATRIN measures a neutrino mass in contradiction to the cosmological neutrino mass bound, this would have exciting consequences for neutrino physics and the most promising cosmological model 
would probably be provided by scenarios that alter the momentum distribution of cosmic relic neutrinos. Conversely, an improved upper limit on the electron neutrino mass from KATRIN 
would increase our confidence in the thermal history of the early Universe.
 
\section*{Acknowledgements}
We acknowledge financial support by MEC and FEDER (EC) Grants No. SEV-2014-0398, FIS2015-2245-EXP, FPA2014-54459 and the Generalitat Valenciana under Grant No. PROME-TEOII/2013/017 (IMO and GB), by the Deutsche Forschungemeinschaft (DFG, German Research Foundation) Research Training Group 1620 `Models of Gravity' (SK and DJS) and Project number 315477589 
TRR 211 (IMO and DJS) and by EU Networks FP10
ITN ELUSIVES (H2020-MSCA-ITN-2015-674896) and INVISIBLES-PLUS (H2020-MSCA-
RISE-2015-690575) (JS and GB), and MINECO grant FPA2016-76005- C2-1-P and Maria de
Maetzu grant MDM-2014-0367 of ICCUB (JS). Numerical calculations have been performed using the computing resources from the Tirant cluster of University of Valencia. 

\appendix
\section{Orthogonal Polynomials}
\label{AppendixA} 

\begin{figure}
\centering
\includegraphics[width=0.65\textwidth]{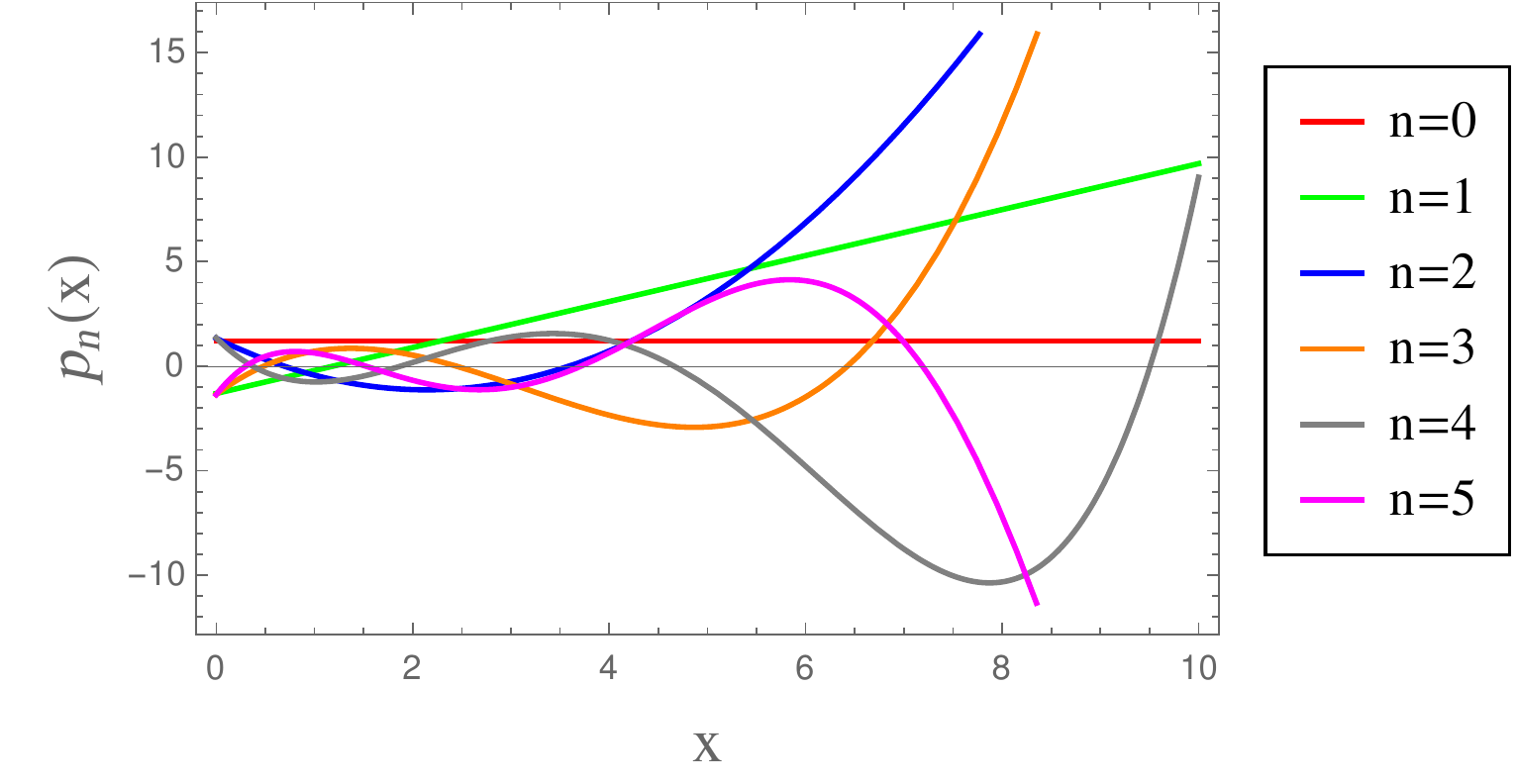}
\caption{The first 6 polynomials that fulfill the orthonormality relation \eqref{eq:orthonormal}. Analytical expressions are given in eq. \eqref{eq:p0}-\eqref{eq:p5}.}
\label{fig:polynomials}
\end{figure}

\allowdisplaybreaks
The first six polynomials fulfilling the orthonormality relation \eqref{eq:orthonormal} are given by
\begin{align}
p_0(x) &=1.201122, \label{eq:p0} \\
p_1(x) &=1.099518 x-1.304654, \label{eq:p1} \\
p_2(x) &= 0.537034 x^2-2.300076 x+1.332213, \label{eq:p2} \\
p_3(x) &=0.176907x^3-1.654993x^2+3.540416x-1.346042, \label{eq:p3} \\
p_4(x) &= 0.043900 x^4 - 0.720629 x^3 + 3.373534 x^2 - 4.804106 x + 1.354667, \label{eq:p4} \\
p_5(x) &= 0.008734 x^5 - 0.222447 x^4 + 1.827305 x^3 - 5.705037 x^2 + 6.083677 x -1.360699. \label{eq:p5}
\end{align}
An illustration of the polynomials \eqref{eq:p0}-\eqref{eq:p5} can be found in figure \ref{fig:polynomials}. In the following, we however only focus on the first three polynomials.

As discussed in section \ref{subsec:Parameterization of relic neutrino distribution}, the expansion in those polynomials \eqref{eq:expansion2} has the disadvantage of evolving negative values for the distribution function for certain choices of parameter combinations $\lbrace F_1,F_2 \rbrace$. This unphysical feature is caused by the orthogonal polynomials themselves having negative values in certain ranges of comoving momentum. In particular,
the term in brackets in equation \eqref{eq:expansion2},
\begin{equation}
p_0(x) + F_1 p_1(x) + F_2 p_2(x),
\label{eq:bracket_term}
\end{equation}
can become negative.

 Let us first study the scenario of $\lbrace F_1 \neq 0, F_2=0 \rbrace$. The $p_1(x)$ polynomial \eqref{eq:p1} has its global minimum at $x=0$. That implies that as long as we have $F_1 < - p_0(0)/p_1(0) \approx 0.92$, equation \eqref{eq:bracket_term} and the distribution function \eqref{eq:expansion2} are positive for all $x$. The lower limit (i.e. $F_1 >0$) is derived by demanding the correct asymptotic behaviour of $f_{\nu}(x)$ \eqref{eq:expansion2} for $x \rightarrow \infty$.

 For $\lbrace F_1=0,F_2 \neq 0 \rbrace$ the same argument gives $F_2 < - p_0(x_0)/p_2(x_0) \approx 1.06$, where $x_0=2.141462$ is the global minimum of $p_2(x)$. For mixed states, i.e. $\lbrace F_1 \neq 0, F_2 \neq 0 \rbrace$, the condition of a positive distribution function for $F_2$ depends on $F_1$ (and the one for $F_1$ on $F_2$). On general grounds (particularly for an MCMC analysis), such  conditions are less convenient to handle. For the sake of simplicity, we therefore hold on the conditions \eqref{eq:conditions}, even though we thereby omit some of the possible parameter space of $\lbrace F_1, F_2 \rbrace$. By restricting our considerations to the parameter ranges \eqref{eq:conditions}, we however still cover a large variety of distribution functions. 
 
\bibliographystyle{utcaps}
\bibliography{Literature}

%%%%%%%%%%%%%%%%%%%%%%%%%%%%%%%%%%%%%%%%%%%

%%%%%%%%%%%%%%%%%%%%%%%%%%%%%%%%%%%%%%%%%%%%%%%%%%%%%%%%%%%%%%%%%%%%%%%%%%%%%%%%%%%%%

\end{document}